\documentclass{article}
\usepackage{amsmath,url,subfigure,graphicx}
\newcommand{\Pro}{\ensuremath{\mathbf{Pr}}}
\usepackage{fancyhdr}

\usepackage{hyperref}
\hypersetup{ %bookmarksopen,
  bookmarksnumbered, 
  pdftitle={A Method for Solving Distributed Service Allocation Problems},
  pdfauthor={Jose M. Vidal},
  pdfsubject={dynamic resource allocation},
  pdfkeywords={},
  pdfview=FitH,
  pdfstartview=FitH 
}

\title{A Method for Solving Distributed Service Allocation
  Problems \footnote{This work has been funded by Darpa under
    contract F30602-99-2-0513.}}
\date{April 17, 2003}
\author{Jos\'{e} M. Vidal \\
Computer Science and Engineering\\
University of South Carolina\\
Columbia, SC 29208\\
\url{vidal@sc.edu}}

\begin{document}
\maketitle

\maketitle
\begin{abstract}
  We present a method for solving service allocation problems in which
  a set of services must be allocated to a set of agents so as to
  maximize a global utility. The method is completely distributed so
  it can scale to any number of services without degradation. We first
  formalize the service allocation problem and then present a simple
  hill-climbing, a global hill-climbing, and a bidding-protocol
  algorithm for solving it. We analyze the expected performance of
  these algorithms as a function of various problem parameters such as
  the branching factor and the number of agents.  Finally, we use the
  sensor allocation problem, an instance of a service allocation
  problem, to show the bidding protocol at work. The simulations also
  show that phase transition on the expected quality of the solution
  exists as the amount of communication between agents increases.
\end{abstract}

\thispagestyle{fancy}
\lhead{}
\chead{Web Intelligence and Agent Systems: An International Journal, 2003.}
%\cfoot{\copyright{} Jos.}

\section{Introduction}
\label{sec:introduction}

The problem of dynamically allocating services to a changing set of
consumers arises in many applications. For example, in an e-commerce
system, the service providers are always trying to determine which
service to provide to whom, and at what price \cite{kephart00a}; in an
automated manufacturing for mass customization scenario, agents must
decide which services will be more popular/profitable \cite{baker99a};
and in a dynamic sensor allocation problem, a set of sensors in a
field must decide which area to cover, if any, while preserving their
resources.

While these problems might not seem related, they are instances of a
more general service allocation problem in which a finite set of
resources must be allocated by a set of autonomous agents so as to
maximize some global measure of utility. A general approach to solving
these types of problems has been used in many successful systems ,
such as \cite{durfee98a} \cite{epstein96a} \cite{wellman96a}
\cite{smith81}. The approach involves three general steps:

\begin{enumerate}
\item Assign each resource that needs to be preserved to an agent
  responsible for managing the resource.
\item Assign each goal of the problem domain to an agent responsible
  for achieving it. Achieving these goals requires the consumption of
  resources.
\item Have each agent take actions so as to maximize its own utility,
  but implement a coordination algorithm that encourages agents to
  take actions that also maximize the global utility.
\end{enumerate}

In this paper we formalize this general approach by casting the
problem as a search in a global \emph{fitness landscape} which is
defined as the sum of the agents' utilities. We show how the choice of
a coordination/communication protocol disseminates information, which
in turn ``smoothes'' the global utility landscape.  This smooth global
utility landscape allows the agents to easily find the global optimum
by simply making selfish decisions to maximize their individual
utility.

We also present experiments that pinpoint the location of a
\emph{phase transition} in the time it takes for the agents to find
the optimal allocation. The transition can be seen when the amount of
communication allowed among agents is manipulated. It exists because
communication allows the agents to align their individual landscapes
with the global landscape. At some amount of communication, the
alignment between these landscapes is good enough to allow the agents
to find the global optimum, but less communication drives the agents
into a random behavior from which the system cannot recuperate.

\subsection{Task Allocation}
\label{sec:task-allocation}

The service allocation problem we discuss in this paper is a superset
of the well known task allocation problem \cite[chapter
5.7]{weiss99a}. A task allocation problem is defined by a set of tasks
that must be allocated among a set of agents. Each agent has a cost
associated with each subset of tasks, which represents the cost the
agent would incur if it had to perform those tasks.  Coordination
protocols are designed to allow agents to trade tasks so that the
globally optimal allocation---the one that minimizes the sum of all
the individual agent costs---is reached as soon as possible. It has
been shown that this globally optimal allocation can reached if the
agents use the contract-net protocol \cite{smith81} with OCSM
contracts \cite{sandholm97a}. These OCSM contracts make it possible
for the system to transition from any allocation to any other
allocation in one step. As such, a simple hill-climbing search is
guaranteed to eventually reach the global optimum.

In this paper we consider the service allocation problem, which is a
superset of the task allocation because it allows for more than one
agent to service a ``task''. The service allocation problem we study
also has the characteristic that every allocation cannot be reached
from every other allocation in one step.

%That is, the global optimum will be reached if the system can get from
%any allocation to any other allocation in one step and the agents only
%take steps that maximize the global utility. This result should be
%intuitive. One can easily surmise that as long as we always move
%uphill and every point on our search space is connected to every other
%point, we are guaranteed to eventually reach the global optimum. In
%this paper we expand this idea and consider the more general service
%allocation problem and the more general situation of landscapes that
%are not fully connected.

\subsection{Service Allocation}
\label{sec:service-allocation}

In a service allocation problem there are a set of services, offered
by service agents, and a set of consumers who use those services. A
server can provide any one of a number of services and some consumers
will benefit from that service without depleting it. A server agent
incurs a cost when providing a service and can choose not to provide
any service.

For example, a server could be an agent that sets up a website with
information about cats. All the consumer agents with interests in cats
will benefit from this service, but those with other interests will
not benefit. Since each server can provide, at most, one service, the
problem is to find the allocation of services that maximizes the sum
of all the agents' utilities, that is, an allocation that maximizes
the global utility.

\subsubsection{Sensor Allocation}
\label{sec:sensor-allocation}

Another instance of the service allocation problem is the sensor
allocation problem, which we will use as an example throughout this
paper. In the sensor allocation problem we have a number of sensors
placed in fixed positions in a two-dimensional space.  Each sensor has
a limited viewing angle and distance but can point in any one of a
number of directions. For example, a sensor might have a viewing angle
of 120 degrees, viewing distance of 3 feet, and be able to look in
three directions, each one 120 degrees apart from the others. That is,
it can ``look'' in any one of three directions. On each direction it
can see everything that is in the 120 degree and 3 feet long view
cone. Each time a sensor looks in a particular direction is uses
energy.

There are also targets that move around in the field. The goal is for
the sensors to detect and track all the targets in the field. However,
in order to determine the location of a target, two or more sensors
have to look at it at the same time. We also wish to minimize the
amount of energy spent by the sensors.

We consider the sensor agents as being able to provide three services,
one for each sector, but only one at a time. We consider the target
agents as consuming the services of the sensors.

\section{A Formal Model for Service Allocation}
\label{sec:model}

We define a service allocation problem $SA$ as a tuple $SA = \{C, S\}$
where $C$ is the set of \textbf{consumer} agents $C = \{c_1, c_2,
\ldots, c_{|C|}\}$, and $c_i$ has only one possible state, $c_i = 0$.
The set of \textbf{service} agents is $S = \{s_1, s_2, \ldots,
s_{|S|}\}$ and the value of $s_i$ is the value of that service. For
the sensor domain in which a sensor can observe any one of three
120-degree sectors or be turned off, we have $s_i \in \{0, 1, 2,
\mbox{off}\}$. An \textbf{allocation} is an assignment of states to
the services (since the consumers have only one possible state we can
ignore them). A particular allocation is denoted by $a = \{s_1, s_2,
\ldots, s_{|S|}\}$, where the $s_i$ have some value taken from the
domain of service states, and $a \in A$, where $A$ is the set of all
possible allocations. That is, an allocation tells us the state of all
agents (since consumers have only one state they can be omitted).

Each agent also has a \textbf{utility} function. The utility that an
agent receives depends on the current allocation $a$, where we let
$a(s)$ be the state of service agent $s$ under $a$. The agent's
utilities will depend on their state and the state of other agents.
For example, in the sensor problem we define the utility of sensor $s$
as $U_{s}(a)$, where
\begin{equation}
  \label{eq:1}
  U_{s}(a) =
  \left\{
      \begin{array}{ll}
        0 &  \mbox{if $a(s) =$ off} \\
        -K_1 & \mbox{otherwise.} \\
      \end{array}
    \right.
\end{equation}
That is, a sensor receives no utility when it is off and must pay a
penalty of $-K_1$ when it is running.

The targets are the consumers, and each target's utility is defined as
\begin{equation}
  \label{eq:2}
  U_{c}(a) = 
  \left\{
    \begin{array}{ll}
      0 & \mbox{if $f_{c}(a) = 0$} \\
      K_2 & \mbox{if $f_{c}(a)=1$} \\
      K_2 + n - 2 & \mbox{if $f_{c}(a)=n$}
    \end{array}
    \right.
\end{equation}
where
\begin{equation}
  \label{eq:3}
  f_{c}(a) = \mbox{number of sensors $s$ that see $c$ given
    their state $a(c)$.}
\end{equation}

Finally, given the individual agent utilities, we define the
\textbf{global utility} $GU(a)$ as the sum of the individual agents'
utilities:
\begin{equation}
  \label{eq:4}
  GU(a) = \sum_{c \in C} U_c(a) + \sum_{s \in S} U_s(a).
\end{equation}

The service allocation problem is to find the allocation $a$ that
maximizes $GU(a)$. In the sensor problem, there are $4^{|S|}$ possible
allocations, which would make a simple generate-and-test approach take
exponential amounts of time. We wish to find the global optimum much
faster than that.

%Finally, we note that the way we have formulated this problem enables
%it to be considered a special case of an NK model of fitness
%landscapes \cite[chapter 8]{at:home:univer}, with $N$ as the number of
%agents and $K$ varying depending on the number of agents on whose
%states an agent's utility depends. This mapping will allow us to apply
%results from $NK$-landscape research to the service allocation
%problem.

%The NK model was designed
%to model the statistical structure of various fitness landscapes,
%ranging from a highly correlated one ($K=0$), to a completely random
%one ($K=N-1$), with multipeaked ruggedness in between.  The correlated
%landscape corresponds to a single big ``mountain'' such that from
%every single starting point we can reach the global maximum by simply
%taking a step upward at every turn. In the random landscape, the
%utility of every point is uncorrelated to the utility of every other
%point, including its neighbors.

% Under the NK model, an agent's utility depends on
%the state of K other agents, and this dependence can be an arbitrary
%function.

%Research on NK landscapes \cite[chapter 10]{at:home:univer} has shown
%that in rugged landscapes the global maximum can be found if agents try
%to maximize the utility of their neighborhood or ``patch,'' rather than
%their own utility. The patch sizes that produce a system dynamic at the
%edge of chaos are the ones that arrive at the global maximum the
%fastest.

\subsection{Search Algorithms}
\label{sec:search-best-alloc}

Our goal is to design an interaction protocol whereby an allocation
$a$ that maximizes the global utility $GU(a)$ is reached in a small
number of steps. In each step of our protocol one of the agents will
change its state or send a message to another agent. The messages
might contain the state or utilities of other agents. We assume that
the agents do not have direct access to the other agents' states or
utility values.

The simplest algorithm we can envision involves having each consumer,
at each time, changing the state of a randomly chosen service agent so
as to increase the consumer's own utility.  That is, a consumer $c$
will change the current allocation $a$ into $a'$ by changing the state
of some sensor $s$ such that $U_c(a') > U_c(a)$. If the sensor's state
cannot be changed so as to increase the utility, then the consumer
does nothing. In the sensor domain this amounts to a target picking a
sensor and changing its state so that the sensor can see the target.
We refer to this algorithm as \textbf{individual hill-climbing}.  

The individual hill-climbing algorithm is simple to implement and the
only communication needed is between the consumer and the chose
server.  This simple algorithm makes every consumer agent increase its
individual utility at each turn. However, the new allocation $a'$
might result in a lower global utility, since $a'$ might reduce the
utility of several other agents. Therefore, it does not guarantee that
an optimal allocation will be eventually reached.

Another approach is for each agent to change state so as to increase
the global utility. We call this a \textbf{global hill-climbing}
algorithm. In order to implement this algorithm, an agent would need
to know how the proposed state change affects the global utility as
well as the states of all the other agents. That is, it would need to
be able to determine $GU(a')$ which requires it to know the state of
all the agents in $a'$ as well as the utility functions of every other
agent, as per the definition of global utility \eqref{eq:4}. In order
for an agent to know the state of others, it would need to somehow
communicate with all other agents. If the system implements a global
broadcasting method then we would need for each agent to broadcast its
state at each time. If the system uses more specialized communications
such as point-to-point, limited broadcasting, etc., then more messages
will be needed.

Any protocol that implements the global hill-climbing algorithm will
reach a locally optimal allocation in the global utility. This is
because it is always true that, for a new allocation $a'$ and old
allocation $a$, $GU(a') \geq GU(a)$. Whether or not this local optimum
is also a global optimum will depend on the ruggedness of the global
utility landscape. That is, if it consists of one smooth peak then it
is likely that any local optimum is the global optimum. On the other
hand, if the landscape is very rugged then there are likely many local
peaks.  Studies in NK landscapes \cite{origin:order} tell us that
smoother landscapes result when an agent's utility depends on the
state of smaller number of other agents.
% [[[already said this?]]]

Global hill-climbing is better than individual hill-climbing since it
guarantees that we will find a local optima. However, it requires
agents to know each others' utility function and to constantly
communicate their state. Such large amount of communication is often
undesirable in multiagent systems. We need a better way to find the
global optimum.

One way of correlating the individual landscapes to the global utility
landscape is with the use of a \textbf{bidding protocol} in which each
consumer agent tells each service the marginal utility the consumer
would receive if the service switched its state to so as to maximize
the consumer's utility. The service agent can then choose to provide
the service with the highest aggregate demand. Since the service is
picking the value that maximizes the utility of everyone involved (all
the consumers and the service) without decreasing the utility of
anyone else (the other services) this protocol is guaranteed to never
decrease the global utility. This bidding protocol is a simplified
version of the contract-net \cite{smith81} protocol in that it does
not require contractors to send requests for bids.

However, in order for a consumer to determine the marginal utility it
will receive from one sensor changing state, it still needs to know
the state of all the other sensors. This means that a complete
implementation of this protocol will still require a lot of
communication (namely, the same amount as in global hill-climbing).
We can reduce this number of messages by allowing agents to
communicate with only a subset of the other agents and making their
decisions based on only this subset of information.  That is, instead
of all services telling each consumer their state, a consumer could
receive state information from only a subset of the services and make
its decision based on this (assuming that the services chosen are
representative of the whole).  This strategy shows a lot of promise
but its performance can only be evaluated on an instance-by-instance
basis.  We explore this strategy experimentally in
Section~\ref{sec:simulations} using the sensor domain.

\subsubsection{Theoretical Time Bounds of Global Hill-Climbing}
\label{sec:time-bounds-global}

Since we now know that global hill-climbing will always reach a local
optimum, the next questions we must answer are:
\begin{enumerate}
\item How many local optima are there?
\item What is the probability that a local optimum is the global
  optimum?
\item How long does it take, on average, to reach a local optimum?
\end{enumerate}

Let $a$ be the current allocation and $a'$ be a neighboring
allocation. We know that $a$ is a local optimum if
\begin{equation}
  \label{eq:5}
  \forall_{a' \in N(a)} GU(a) > GU(a')
\end{equation}
where
\begin{equation}
  \label{eq:13}
  N(a) = \{ x \, | \, x \mbox{ is a Neighbor of } a \}.
\end{equation}
We define a \textit{Neighbor} allocation as an allocation where one,
and only one, agent has a different state.

The probability that some allocation $a$ is a local optimum is simply
the probability that ~\eqref{eq:5} is true. If the utility of all
pairs of neighbors is not correlated, then this probability is
\begin{equation}
  \label{eq:7}
  \Pro[\forall_{a' \in N(a)} GU(a) > GU(a')]
  = \Pro[GU(a) > GU(a')]^b, 
\end{equation}
where $b$ is the \textbf{branching factor}. In the sensor problem $b =
3 \cdot |S|$ where $S$ is the set of all sensors. That is, since each
sensor can be in any of four states it will have three neighbors from
each state. In some systems it is safe to assume that the global
utilities of $a$'s neighbors are independent. However, most systems
show some degree of correlation.

Now we need to calculate the $\Pro[GU(a) > GU(a')]$, that is, the
probability that some allocation $a$ has a greater global utility that
its neighbor $a'$, for all $a$ and $a'$. This could be calculated via
an exhaustive enumeration of all possible allocations. However, often
we can find the expected value of this probability.

For example, in the sensor problem each sensor has four possible
states. If a sensor changes its state from sector $x$ to sector $y$
the utility of the target agents covered by $x$ will decrease while
the utility of those in $y$ will increase. If we assume that, on
average, the targets are evenly spaced on the field, then the global
utilities for both of these are expected to be the same. That is, the
expected probability that the global utility of one allocation is
bigger than the other is $1/2$.

If, on the other hand, a sensor changes state from ``off'' to a
sector, or from a sector to ``off,'' the global utility is expected to
decrease and increase, respectively. However, there are an equal
number of opportunities to go from ``off'' to ``on'' and vice-versa.
Therefore, we can also expect that for these cases the probability
that the global utility of one allocation is bigger than the other is
$1/2$.

Based on these approximations, we can declare that for the sensor
problem
\begin{equation}
  \label{eq:6}
  \Pro[\forall_{a' \in N(a)} GU(a) > GU(a')]
  =\frac{1}{2^b} = \lambda.
\end{equation}

If we assume an even distribution of local optima, the total number of
local optima is simply the product of the total number of allocations
times the probability that each one is a local optimum. That is,
\begin{equation}
  \label{eq:8}
  \mbox{Total number of local optima} = \lambda |A|
\end{equation}

For the sensor problem, $\lambda = 1/2^b$, $b = 3 \cdot |S|$ and $|A|
= b^{|S|}$, so the expected number of local optima is $b^{|S|} /
2^{3|S|}$.

\begin{equation}
  \label{eq:9}
  \Pro[\mbox{a local optimum is also global}] = \frac{1}{\lambda |A|} = 
  \frac{1}{2^b}.
\end{equation}

We can find the expected time the algorithm will take to reach a local
optimum by determining the maximum number of steps from every
allocation to the nearest local optimum. This gives us an upper bound
on the number of steps needed to reach the nearest local optimum using
global hill-climbing. Notice that, under either individual
hill-climbing or the bidding protocol it is possible that the local
optimum is not reached, or is reached after more steps, since these
algorithms can take steps that lower the global utility.

In order to find the expected number of steps to reach a local
optimum, we start at any one of the local optima and then traverse all
possible links at each depth $d$ until all possible allocations have
been visited. This occurs when
\begin{equation}
  \label{eq:10}
  \lambda \cdot |A|\cdot b^d > |A|.
\end{equation}
Solving for $d$, and remembering that $\lambda = 1/2^b$, we get
\begin{equation}
  \label{eq:11}
  d > b \log_b 2.
\end{equation}

The expected worst-case distance from any point to the nearest local
optimum is, therefore, $b \log_b 2$ (this number only makes sense for
$b \geq 2$ since smaller number of neighbors do not form a searchable
space). That is, the number of steps to reach the nearest local optima
in the sensor domain is proportional to the branching factor $b$,
which is equal to $3 \cdot |S|$. We can expect search time to increase
linearly with the number of sensors in the field.

\section{Simulations}
\label{sec:simulations}

While the theoretical results above give us some bounds on the number
of iterations before the system is expected to converge to a local
optimum, the bounds are rather loose and do not tell us much about the
dynamics of the executing system. Also, we cannot show mathematically
how changes in the amount of communication change the search.
Therefore, we have implemented a service allocation simulator to
answer these questions. It simulates the sensor allocation domain
described in the introduction.

The simulator is written in Java and the source code is available upon
request.  It gathers and analyzes data from any desired number of
runs. The program can analyze the behavior of any number of target and
sensor agents on a two-dimensional space, and the agents can be given
any desired utility function.  The program is limited to static
targets. That is, it only considers the one-shot service allocation
problem. Each new allocation is completely independent of any previous
one.

In the tests we performed, each run has seven sensors and seven
targets, all of which are randomly placed on a two-dimensional grid.
Each sensor can only point in one of three directions or sectors.
These three sectors are the same for all sensors (specifically, the
first sector is from 0 to 120 degrees, the second one from 120 to 240,
and the third one from 240 to 360). All the sensors use the same
utility function which is given by ~\eqref{eq:1}, while the targets
use ~\eqref{eq:2}. After a sensor agent receives all the bids it
chooses the sector that has the heighest aggregate demand, as
described by the bidding protocol in
Section~\ref{sec:search-best-alloc}.

During a run, each of the targets periodically sends a bid to a number
of sensors asking them to turn to the sector that faces the target.
We set the bid amount to a fixed number for these tests.
Periodically, the sensors count the number of bids they have received
for each sector and turn their detector (such as a radar) to face the
sector with the highest aggregate demand. We assume that neither the
targets nor the sensors can form coalitions.

We vary the number of sensors to which the targets send their bids in
order to explore the quality of the solution that the system converges
upon as the amount of communication changes. For example, at one
extreme if the all the targets send their bids to all the sensors,
then the sensors would always set their sector to be the one with the
most targets. This particular service allocation should, usually, be
the best. However, it might not always be the optimal solution. For
example, if seven targets are clustered together and the eighth is on
another part of the field, it would be better if six sensor agents
pointed towards the cluster of targets while the remaining two sensor
agents pointed towards the stray target rather than having all sensor
agents point towards the cluster of targets. At the other extreme, if
all the targets send their bids to only one sensor then they will
minimize communications but then the sensors will point to the sector
from which they received a message---an allocation which is likely to
be suboptimal.

These simulations explore the ruggedness of the system's global
utility landscape and the dynamics of the agents' exploration of this
landscape. If the agents were to always converge on a local
(non-global) optimum then we would deduce that this problem domain has
a very rugged utility landscape. On the other hand, if they usually
manage to reach the global optimum then we could deduce a smooth
utility landscape.

\begin{figure*}[tb]
  \begin{center}
    \mbox{\subfigure{\label{fig:k=1}
        \includegraphics[width=.5\columnwidth]{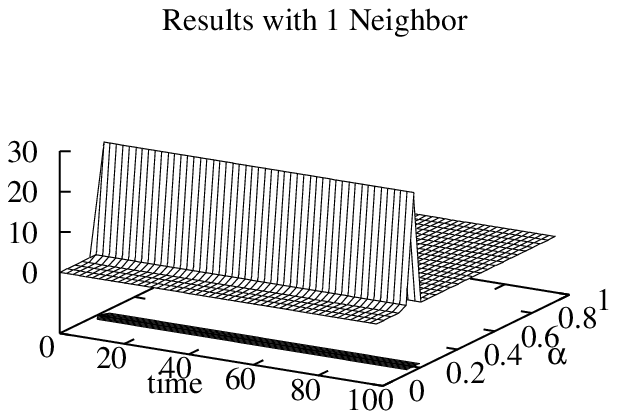}
        }
      \subfigure{\label{fig:k=3}
        \includegraphics[width=.5\columnwidth]{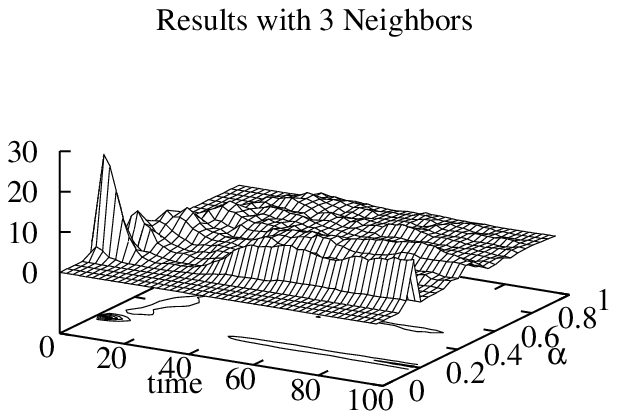}}
      } \\
    \mbox{\subfigure{\label{fig:k=5}
        \includegraphics[width=.5\columnwidth]{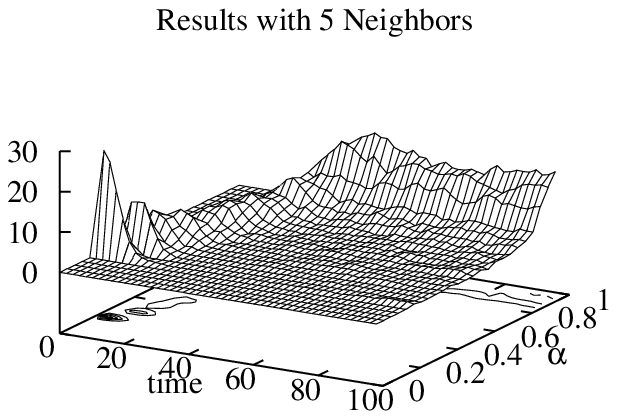}
        }
      \subfigure{\label{fig:k=7}
        \includegraphics[width=.5\columnwidth]{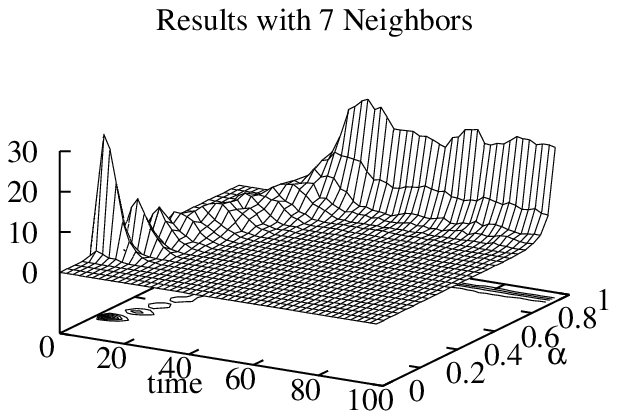}
        }} \\
    \caption{The z-axis on each figure represents the number of runs,
      out of 100, which had the particular $\alpha$ ratio at each
      particular time. $\alpha = 1$ means the run is at the global
      optimum. The optimum is reached more often in the cases with more 
      communication.}
    \label{fig:results}
  \end{center}
\end{figure*}

\begin{figure}[tbhp]
  \begin{center}
    \includegraphics[width=\columnwidth]{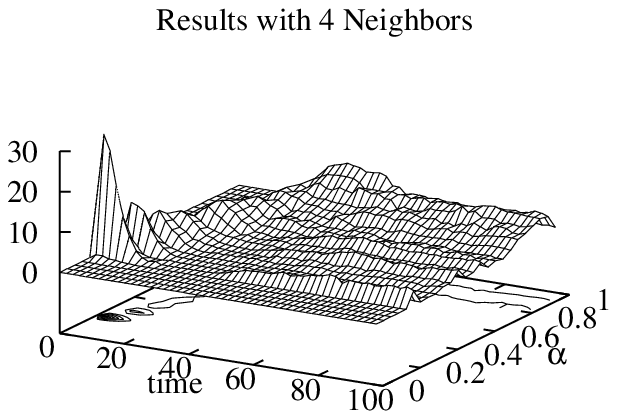}
    \caption{The transitional case occurs when the target communicates
    with four sensors.}
    \label{fig:trans}
  \end{center}
\end{figure}

\section{Test Results}
\label{sec:test-results}

In each of our tests we set the number of agents that each target will
send its bid to, that is, the number of neighbors, to a fixed number.
Given this fixed number of neighbors, we then generated 100 random
placements of agents on the field and ran our bidding algorithm 10
times on each of those placements. Finally, we plotted the average
solution quality, over the 10 runs, as a function of time for each of
the 100 different placements. The solution quality is given by the
ratio
\begin{equation}
  \label{eq:12}
    \alpha = \frac{\mbox{Current Utility}}{\mbox{Globally Optimal Utility}},
\end{equation}
so if $\alpha = 1$, then it means that the run has reached the global
optimum. Since the number of agents is small, we were able to
calculate the global optimum using a brute-force method. Specifically,
there are $3^7 = 2187$ possible configurations times $100$ random
placements leads to $218700$ combinations that we had to check for
each run in order to find the global optimum using brute-force. Using
more than $7$ sensors made the test take too long. Notice, however,
that our algorithm is much faster than this brute-force search which
we perform only to confirm that our search does find the global
optimum.

In our tests there were always seven target agents and seven sensor
agents. We varied the number of neighbors from 1 to 7.  If the target
can only communicate with one other sensor, the sensors will likely
have very little information for making their decision, while if all
targets communicate with all seven sensors, then each sensor will
generally be able to point to the sector with the most targets.
However, because these decisions are made in an asynchronous manner,
it is possible that some sensor will sometimes not receive all the
bids before it has to make a decision. The targets always send their
bids to the sensors that are closest to them.

The results from our experiments are shown in Figure~\ref{fig:results}
where we can see that there is a transition in the system's
performance as the number of neighbors goes from three to five. That
is, if the targets only send their bids to three sensors then it is
almost certain that the system will stay in a configuration that has a
very low global utility. However, if the targets send their bids to
five sensors, then it is almost guaranteed ($98\%$ of the time) that
the system will reach the globally optimal allocation. This is a huge
difference in terms of the performance. We also notice in
Figure~\ref{fig:trans} that for four neighbors there is a fairly even
distribution in the utility of the final allocation.

\section{Related Work}
\label{sec:related-work}

There is ongoing work in the field of complexity that attempts to
study they dynamics of complex adaptive systems \cite{origin:order}. Our
approach is based on ideas borrowed from the use of NK landscapes for
the analysis of co-evolving systems. As such, we are using some of the 
results from that field. However, complexity theory is more concerned
with explaining the dynamic behavior of existing systems, while we are 
more concerned with the engineering of multiagent systems for
distributed service allocation.

The Collective Intelligence (COIN) framework \cite{wolpert99a} shares
many of the same goals of our research. They start with a global
utility function from which they derive the rewards functions for each
agent. The agents are assumed to use some form of reinforcement
learning. They show that the global utility is maximized when using
their prescribed reward functions. They do not, however, consider how
agent communication might affect the individual agent's utility
landscape.

The task allocation problem has been studied in
\cite{rules:of:encounter}, but the service allocation problem we
present in this paper has received very little attention.  There is
also work being done on the analysis of the dynamics of multiagent
systems for other domains such as e-commerce \cite{kephart00a} and
automated manufacturing \cite{parunak97b}. It is possible that
extensions to our approach will shed some light into the dynamics of
these domains.

\section{Conclusions}
\label{sec:conclusions}

We have formalized the service allocation problem and examined a
general approach to solving problems of this type. The approach
involves the use of utility-maximizing agents that represent the
resources and the services. A simple form of bidding is used for
communication. An analysis of this approach reveals that it implements
a form of distributed hill-climbing, where each agent climbs its own
utility landscape and not the global utility landscape. However, we
showed that increasing the amount of communication among the agents
forces each individual agent's landscape to become increasingly
correlated to the global landscape.

These theoretical results were then verified in our implementation of
a sensor allocation problem---an instance of a service allocation
problem. Furthermore, the simulations allowed us to determine the
location of a phase transition in the amount of communication needed
for the system to consistently arrive at the globally optimal service
allocation. 

More generally, we have shown how a service allocation problem can be
viewed as a distributed search by multiple agents over multiple
landscapes. We also showed how the correlation between the global
utility landscape and the individual agent's utility landscape depends
on the amount and type of inter-agent communication.  Specifically, we
have shown that increased communications leads to a higher correlation
between the global and individual utility landscapes, which increases
the probability that the global optimum will be reached. Of course, the
success of the search still depends on the connectivity of the search
space, which will vary from domain to domain. We expect that our
general approach can be applied to the design of any multiagent
systems whose desired behavior is given by a global utility function
but whose agents must act selfishly.

Our future work includes the study of how the system will behave under
perturbations. For example, as the target moves it perturbs the
current allocation and the global optimum might change. We also hope
to characterize the local to global utility function correlation for
different service allocation problems and the expected time to find
the global optimum under various amounts of communication.

%For example, once the sensors have reached a globally optimal
%allocation, the targets might move and the sensors will have to
%re-allocate themselves. We expect that for small movements the sensors
%will very quickly find the new globally optimal allocation. We are
%also interested in extending the utility values to include time.  This
%way, for example, we could require a sensor to flip back and forth
%between two settings in order to achieve the best allocation of its
%resources. As it stands, our utility values can depend only on the
%current state of other agents, not on their past states.

%\section{Acknowledgements}
%\label{sec:acknowledgements}

%(Deleted for blind review).

\bibliographystyle{abbrv-jmv} 
\bibliography{../../library,../../vidal}

\begin{thebibliography}{10}

\bibitem{baker99a}
A.~D. Baker, H.~V. Parunak, and K.~Erol.
\newblock \href{http://jmvidal.cse.sc.edu/library/baker99a.pdf}{Agents and the
  internet: Infrastructure for mass customization}.
\newblock {\em {IEEE} Internet Computing}, 3(5):62--69, September-October 1999.

\bibitem{durfee98a}
E.~H. Durfee, T.~Mullen, S.~Park, J.~M. Vidal, and P.~Weinstein.
\newblock \href{http://jmvidal.cse.sc.edu/papers/umdlsms/}{The dynamics of the
  {UMDL} service market society}.
\newblock In M.~Klusch and G.~Wei\ss, editors, {\em Cooperative Information
  Agents {II}}, {LNAI}, pages 55--78. Springer, 1998.

\bibitem{epstein96a}
J.~M. Epstein and R.~L. Axtell.
\newblock \href{http://www.brook.edu/sugarscape}{{\em Growing Artificial
  Societies : Social Science from the Bottom Up}}.
\newblock Brookings Institute, 1996.

\bibitem{origin:order}
S.~Kauffman.
\newblock
  \href{http://www.amazon.com/exec/obidos/ASIN/0195079515/multiagentcom}{{\em
  The Origins of Order: Self-Organization and Selection in Evolution}}.
\newblock Oxford University Pres, 1993.

\bibitem{kephart00a}
J.~O. Kephart, J.~E. Hanson, and A.~R. Greenwald.
\newblock
  \href{http://www.research.ibm.com/infoecon/paps/html/rudin/rudin.html}{Dynam%
ic pricing by software agents}.
\newblock {\em Computer Networks}, 32(6):731--752, 2000.

\bibitem{parunak97b}
H.~V.~D. Parunak.
\newblock \href{http://jmvidal.cse.sc.edu/library/parunak97b.pdf}{``go to the
  ant'': Engineering principles from natural agent systems}.
\newblock {\em Annals of Operation Research}, 75:69--101, 1997.

\bibitem{rules:of:encounter}
J.~S. Rosenschein and G.~Zlotkin.
\newblock
  \href{http://www.amazon.com/exec/obidos/ASIN/0262181592/multiagentcom}{{\em
  Rules of Encounter}}.
\newblock The {MIT} Press, Cambridge, MA, 1994.

\bibitem{sandholm97a}
T.~W. Sandholm.
\newblock Necessary and sufficient contract types for optimal task allocation.
\newblock In {\em Proceedings of the Fourteenth International Joint Conference
  on Artificial Intelligence}, 1997.

\bibitem{smith81}
R.~G. Smith.
\newblock The contract net protocol: High-level communication and control in a
  distributed problem solver.
\newblock {\em {IEEE} Transactions on Computers}, C-29(12):1104--1113, 1981.

\bibitem{weiss99a}
G.~Weiss, editor.
\newblock \href{http://jmvidal.cse.sc.edu/library/WeissBook/}{{\em Multiagent
  Systems: A Modern Approach to Distributed Artificial Intelligence}}.
\newblock {MIT} Press, 1999.

\bibitem{wellman96a}
M.~P. Wellman.
\newblock
  \href{http://jmvidal.cse.sc.edu/library/wellman96a.ps}{Market-oriented
  programming: Some early lessons}.
\newblock In S.~Clearwater, editor, {\em Market-Based Control: A Paradigm for
  Distributed Resource Allocation.} World Scientific, 1996.

\bibitem{wolpert99a}
D.~H. Wolpert and K.~Tumer.
\newblock \href{http://xxx.lanl.gov/abs/cs.LG/9908014}{An introduction to
  collective intelligence}.
\newblock Technical report, ACM Computing Research Repository, 1999.
\newblock cs.LG/9908014.

\end{thebibliography}

\end{document}